\documentclass[a4paper,11pt]{article}
\pdfoutput=1 
\usepackage{graphicx}
\usepackage{jinstpub} 

\title{First results on the performance of the PADME electromagnetic calorimeter}

\author[a,b,1]{G. Piperno,\note{Corresponding author.}}

\affiliation[a]{Dipartimento di Fisica, Sapienza -- Universit\`a di Roma, Roma I-00185, Italy}
\affiliation[b]{INFN -- Sezione di Roma, Roma I-00185, Italy}

\emailAdd{gabriele.piperno@roma1.infn.it}

\abstract{The PADME experiment, hosted at the Laboratori Nazionali di Frascati
in Italy, is dedicated to the dark photon search. It looks for the
reaction $e^{+}\,e^{-}\rightarrow A'\,\gamma$, where $A'$ indicates
the dark photon. The $550\,\text{MeV}$ positrons beam that impinges
on electrons, coming from an active diamond target, allows to scan
$A'$ masses up to $23.7\,\text{MeV}$.
In this context, the segmented electromagnetic calorimeter plays a
fundamental role, since it measures the final photon four-momentum.
It consists of $616$ BGO crystals displaced in a cylindrical shape of $\approx29\,\text{cm}$ radius with
a central square hole of $5$ crystals side, to let Bremsstrahlung
radiation pass, to limit the calorimeter trigger rate. 
Each crystal is read by a HZC XP1911 type B photomultipliers and their
signal is digitised by means of a CAEN V1742 board, taking $1024$ samples
at $1\,\text{GS/s}$.
Here we report in detail the solutions adopted for the calorimeter
together with the results obtained in tests performed on a small prototype
and on single scintillating units.}

\keywords{Calorimeters, Scintillators, Gamma detectors, Detector design and construction technologies and materials}

\collaboration[c]{on behalf of PADME collaboration}

\proceeding{15$^{\text{th}}$ Topical Seminar on Innovative Particle and Radiation Detectors (IPRD19)
\\ 14-17 October 2019 \\ Siena, Italy}

\begin{document}
\maketitle
\flushbottom

\section{Introduction}

As a result of many cosmological and astronomical observations we
know that only $15\%$ of the mass of the Universe is made of
Standard Model (SM) matter. The largest part seems to consist of an
invisible form of matter, called Dark Matter (DM) \cite{Planck 2018}.
Despite all these results, DM nature is still unknown.

A possible solution to the problem of its elusiveness is that it lives
in a separate sector, named Dark Sector, and that it interacts
with the SM only by means of portals. The simplest connection between
our world and the Dark Sector can be built introducing a new U(1) symmetry,
which adds a new vector boson, the Dark Photon (DP) $A'$ \cite{DP1,DP2}.
Even if the SM particles are neutral under this symmetry, the DP can
interact with them via a kinetic mixing with the ordinary photon. Consequently, particles
will acquire an effective charge $\varepsilon q$, where $\varepsilon$
is the $A'$ coupling constant and $q$ is the particle electric charge.

A strong point of this model is that it is very predictive, being
completely described when the constant $\varepsilon$ and the mass
$m_{A'}$ are defined.

Generally the DP search is divided into two classes, depending on
the decay channel type: visible or invisible. The first case occurs
when there are no DM particles with mass $\leq m_{A'}/2$ and the
$A'$ boson is forced to decay into SM particles. On the contrary, when
DM particles with mass $\leq m_{A'}/2$ exist, the $A'$ boson will decay
mostly into this kind of particles and the SM channels are then suppressed
by a factor $\varepsilon^{2}$.

The DP can also solve other existing tensions of the SM. Depending
on the model, it could explain, partially or completely, the discrepancy
between the expected value and the measured one of the muon anomalous
magnetic moment $\left(g-2\right)_{\mu}$ \cite{g-2} and the two
recent observations known as the Beryllium-$8$ and Helium-4 anomaly
\cite{Be-8,He-4}.

In refs. \cite{DP rew1,DP rew2} a more complete panorama of the DP
physics and current research approaches are presented.

\section{The PADME experiment}

The main aim of the Positron Annihilation into Dark Matter Experiment
(PADME), hosted at the Laboratori Nazionali di Frascati, is
the search for an invisibly decaying DP. Placed in the laboratories Beam Test
Facility \cite{BTF}, it exploits the LINAC to accelerate
positrons that then annihilate with electrons coming from a target, to look for the reaction:

\[
e^{+}\,e^{-}\rightarrow A'\,\gamma.
\]

Positrons are accelerated up to $550\,\text{MeV}$, while the electrons
are at rest in a diamond active target. If produced, a long lived
or invisibly decaying $A'$ can't be observed by the experiment. On
the contrary, the $\gamma$ four-momentum can
be measured by means of a granular electromagnetic calorimeter. Consequently,
the $A'$ boson can be observed as missing mass that, knowing the initial
conditions ($\vec{P}_{e^{-}}=\vec{0}\,\text{MeV/c}$ and $\vec{P}_{e^{+}}=(0,0,550)\,\text{MeV/c}$),
can be measured as (the $\underline{P}$ notation indicates the relativistic four-momentum):
\[
M_{miss}^{2}=\left(\underline{P}_{e^{-}}+\underline{P}_{e^{+}}-\underline{P}_{\gamma}\right)^{2}.
\]

Requiring only the coupling to leptons, the PADME DP search results
to be almost model independent.

Figure \ref{fig:PADME detector} represents the scheme of the detector.
Moving from the right to the left of the image, it is composed of
the following elements:

\begin{figure}
\begin{centering}
\includegraphics[scale=0.5]{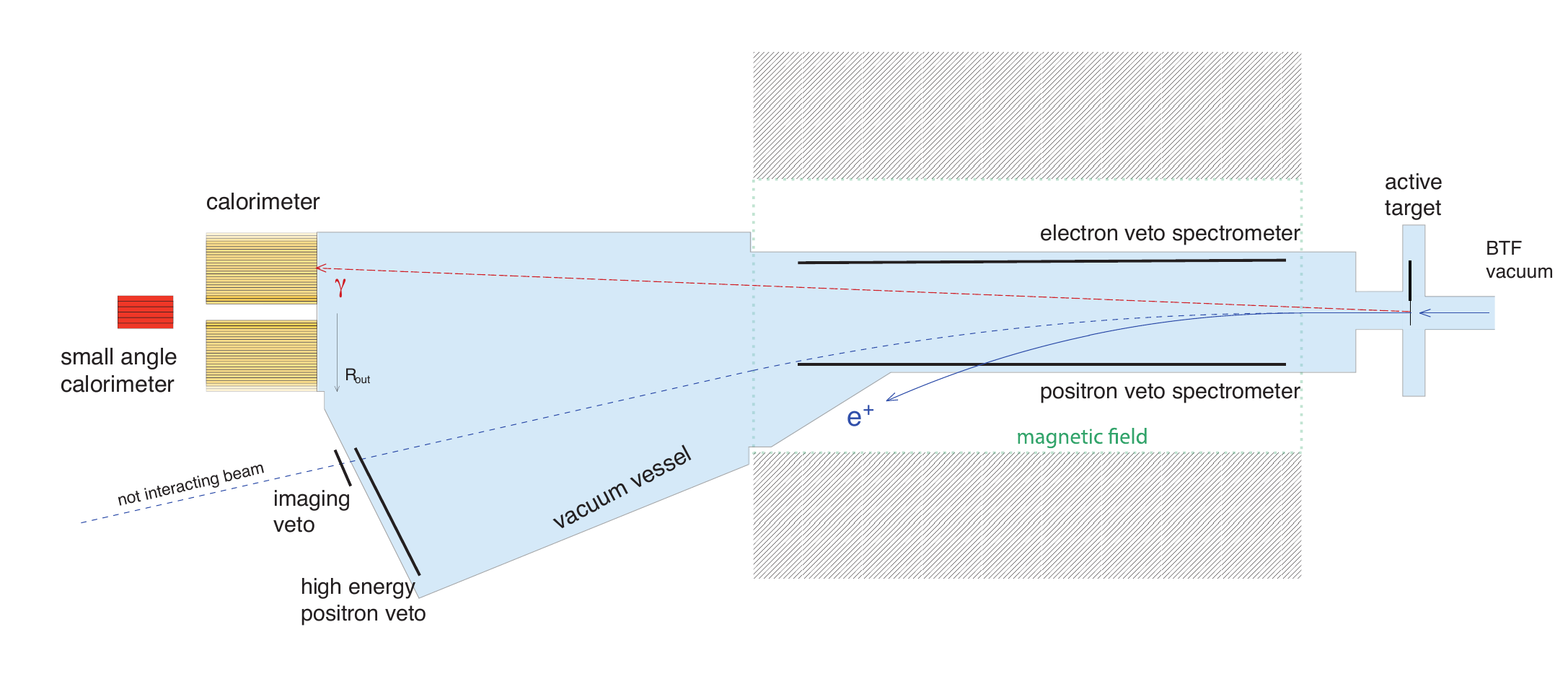}
\par\end{centering}
\caption{\label{fig:PADME detector}PADME detector scheme, with positron beam
arriving from the right: diamond active target, magnetic dipole with
electron/positron vetoes inside and high energy positron veto and
TimePix3 array, both close to the beam exit; at $3.45\,\text{m}$
from the target are placed the electromagnetic calorimeter and, $50\,\text{cm}$ behind it,
the small angle calorimeter.}
\end{figure}

\begin{itemize}
\item The active target consists of a $100\,\mu\text{m}$ thick
and $2\times2\,\text{cm}^{2}$ area polycrystalline diamond, with
$19$ horizontal and $19$ vertical graphitic stripes (only $16$
are connected). It returns average information about the beam intensity
and position (resolution $\approx5\,\text{mm}$). The material is
chosen to reduce the Bremsstrahlung cross section, while the thickness
is a balance between threshold (the minimum number of detectable positrons),
which decreases with thickness, and
the beam multiple scattering, which increases with thickness.
\item The magnetic dipole, a MBP-S dipole $1\,\text{m}$ long with $23\,\text{cm}$
gap and $\approx0.5\,\text{T}$ field, is placed $20\,\text{cm}$
after the target. Its purpose is to bend beam particles sending them
towards the experiment exit, if did not annihilate on the target,
or to the charged particle veto system, if they lost part of their
energy by Bremsstrahlung. 
\item The charged particle veto system is the sum of different components:
two arrays of plastic scintillating bars inside the magnet gap, one
for $e^{+}$ ($90\,\text{cm}$ long), one for $e^{-}$ ($96\,\text{cm}$
long), needed to identify either positrons that lost a large amount
of energy for Bremsstrahlung, or the electrons and positrons produced
in $e^{+}\,e^{-}\rightarrow e^{+}\,e^{-}$ interactions. Another array
($36\,\text{cm}$ long), close to the beam exit, is used to detect
positrons that lost a small amount of energy for Bremsstrahlung. Each
scintillating bar size is $1\times1\times16\,\text{cm}^{3}$.
\item An array of $6\times2$ TimePix3 silicon detectors, for a total size
of $8.45\times2.82\,\text{cm}^{2}$, is positioned at the beam exit
to study bunch divergence and structure, thanks to its spatial ($55\,\mu\text{m}$
pitch) and time ($<0.5\,\text{ns}$) resolutions.
\item On the opposite side of the experiment from the target there is the
Electromagnetic Calorimeter (ECal). It presents a central square opening
that allows the Bremsstrahlung radiation to pass and to be detected
by a faster calorimeter. Additional information on the ECal is given
in the next section.
\item The small angle calorimeter, placed $50\,\text{cm}$ behind
the ECal, has a square shape and is made of $25$ $3\times3\times14\,\text{cm}^{3}$
lead difluoride (PbF$_{2}$) crystals. The readout signal is Cherenkov
light ($\approx3\,\text{ns}$ duration), which makes the detector
able to sustain an event rate up to hundreds of MHz.
\end{itemize}
A signal event, $e^{+}\,e^{-}\rightarrow A'\,\gamma$, would appear
as a single $\gamma$ in the ECal with no hits in the small angle calorimeter or vetoes
and some missing energy, corresponding to the amount taken away by
the $A'$ boson. Given the beam energy of $550\,\text{MeV}$, the maximum
$A'$ mass that can be explored is $23.7\,\text{MeV}$.

\begin{figure}
\begin{centering}
\includegraphics[scale=0.5]{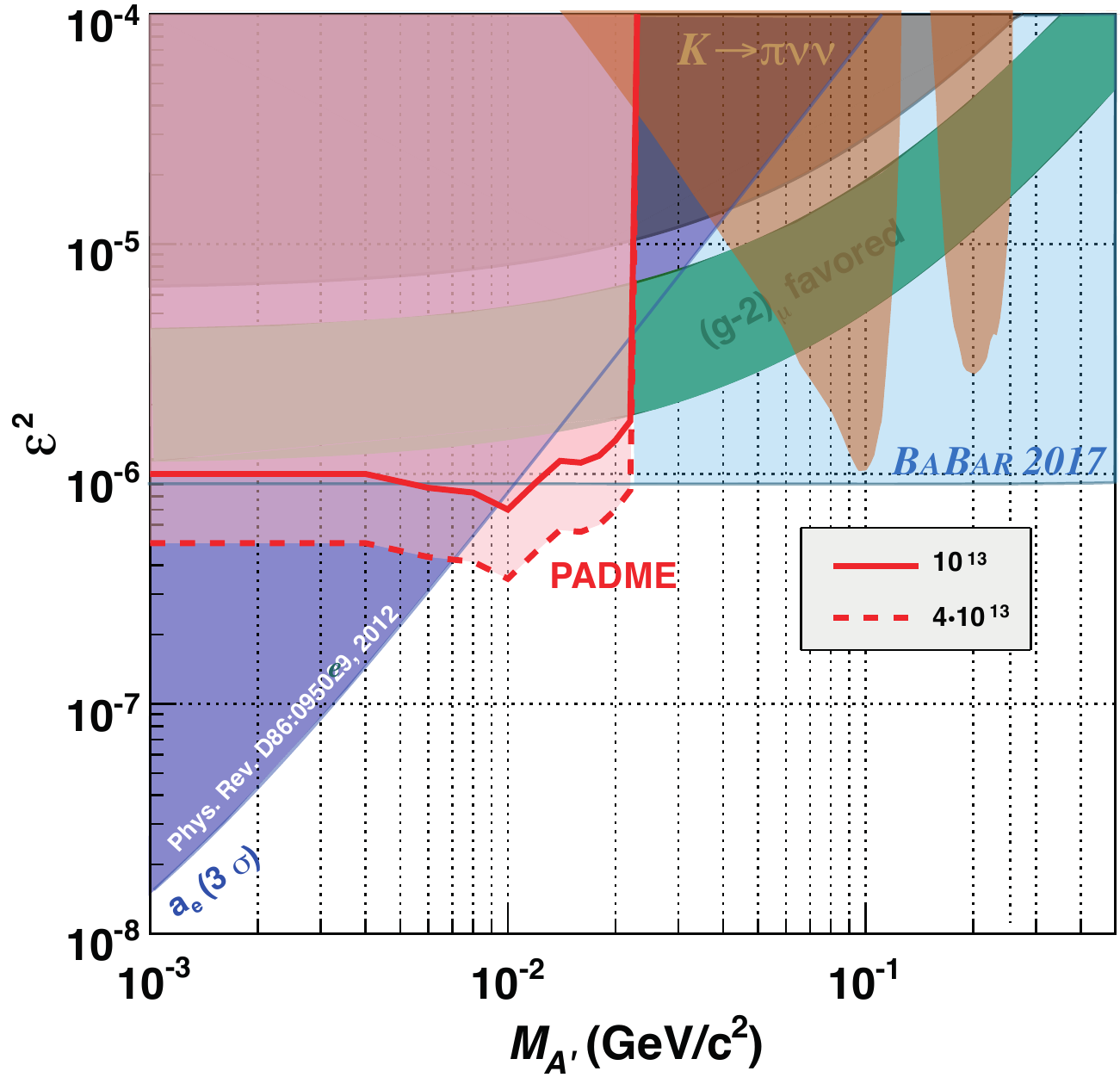}
\par\end{centering}
\caption{\label{fig:PADME sensitivity}PADME $90\%$ C.L. sensitivity to a DP that decays
into DM particles considering two different amount of positrons on
target: $10^{13}$ and $4\cdot10^{13}$.}
\end{figure}

The PADME $90\%$ C.L. sensitivity to an $A'$ boson that decays in the invisible channel is presented
in figure \ref{fig:PADME sensitivity}. Here the result is given assuming
the kinetic mixing model introduced in the previous section and for two different
statistics: $10^{13}$ and $4\cdot10^{13}$ positrons on target.

\section{The PADME electromagnetic calorimeter}

The ECal, together with the active target, is the most important component
of the experiment, given that it measures the four-momentum of the
photon in the final state. The chosen material is BGO (bismuth
germanate) because of its small radiation length ($X_{0}=1.12\,\text{cm}$)
and because the PADME collaboration had the opportunity to reshape
and reuse crystals from L3 electromagnetic calorimeter endcaps \cite{L3}.
It is composed of $616$ BGO scintillating crystals of dimensions
$2.1\times2.1\times23.0\,\text{cm}^{3}$, displaced in an almost cylindrical
shape with a radius of about $29\,\text{cm}$. Given the long decay
time of the BGO scintillation light, $300\,\text{ns}$, the ECal is
not able to sustain high particle rates. Consequently, it has a central
square hole of $5\times5$ crystals to let the Bremsstrahlung radiation,
very intense at small angles, pass and be observed by the faster
small angle calorimeter. With the current distance of $3.45\,\text{m}$ from the active
target, the ECal angular coverage is $[15,84]\,\text{mrad}$.

\begin{figure}
\begin{centering}
\includegraphics[scale=0.5]{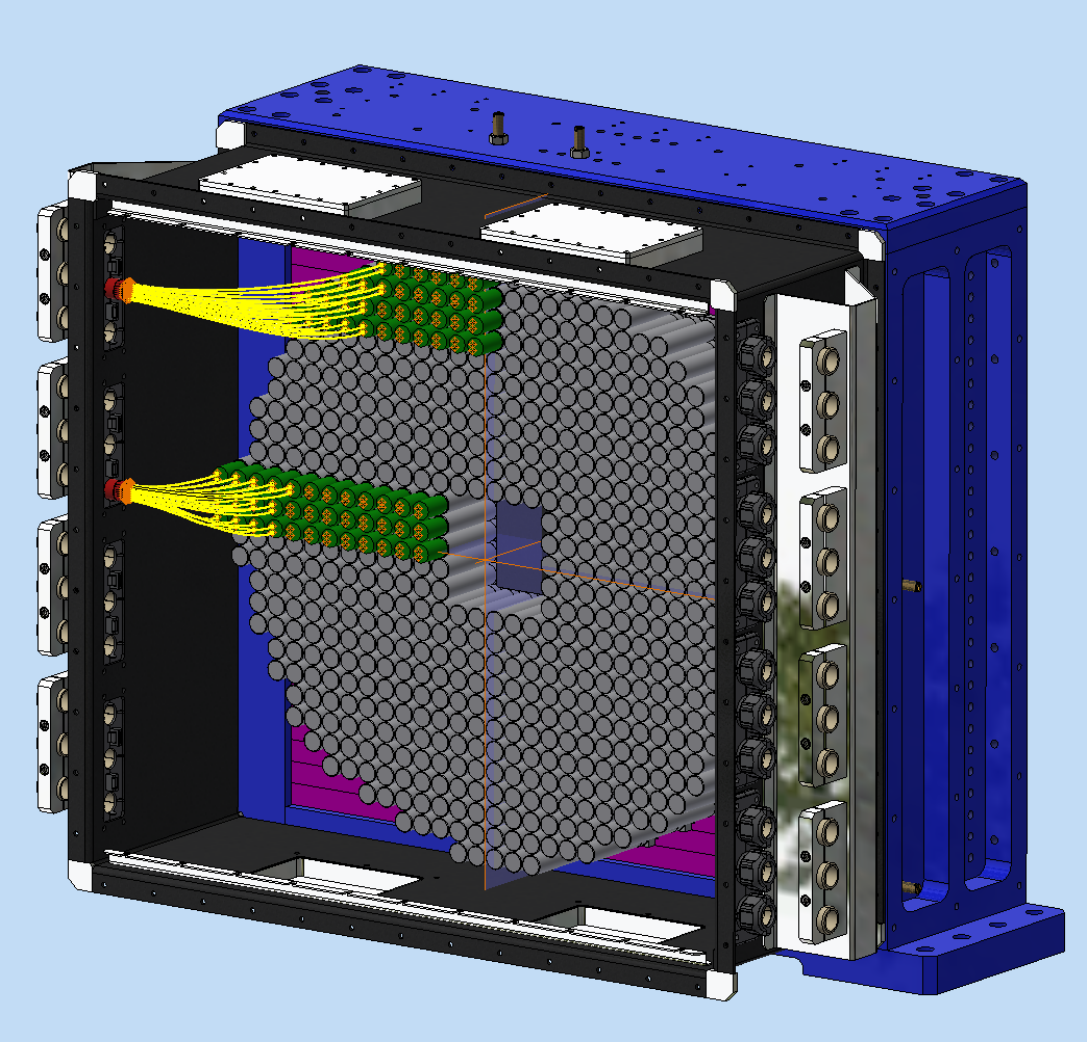}
\par\end{centering}
\caption{\label{fig:PADME ECal}Render of the PADME ECal, with some of the
PMTs (orange) and cables (yellow) drawn.}
\end{figure}

A photomultiplier tube (PMT) is glued to each BGO crystal to form
the so called Scintillating Unit (SU). The selected model is the HZC
XP1911 type B, which fits the crystals' square cross section well,
thanks to the diameter of $1.9\,\text{cm}$, and the BGO emission
spectrum, that has a maximum at $480\,\text{nm}$, 
at which wavelength the quantum efficiency of the PMT(s) is $21\%$ \cite{HZC Photonics}.
A render of the calorimeter is shown in figure \ref{fig:PADME ECal},
in which some of the PMTs (in orange) and signal and HV cables (in
yellow) can be seen.

The ECal does not have a honeycomb structure in which to insert the
various SUs, due to the fact that it would have spoiled the energy
resolution of the calorimeter. Therefore one layer of crystals acts
as basis for the next layer. To reduce the light crosstalk, $50\,\mu\text{m}$
black Tedlar foils are placed between the SUs. These foils are also
used to level out any difference in crystal heights by varying the
number of layers. Since the start of calorimeter operations in October
2018, $4$ SUs have never worked, corresponding to less than the $0.7\%$
of the total.

\begin{figure}
\begin{centering}
\includegraphics[scale=0.3]{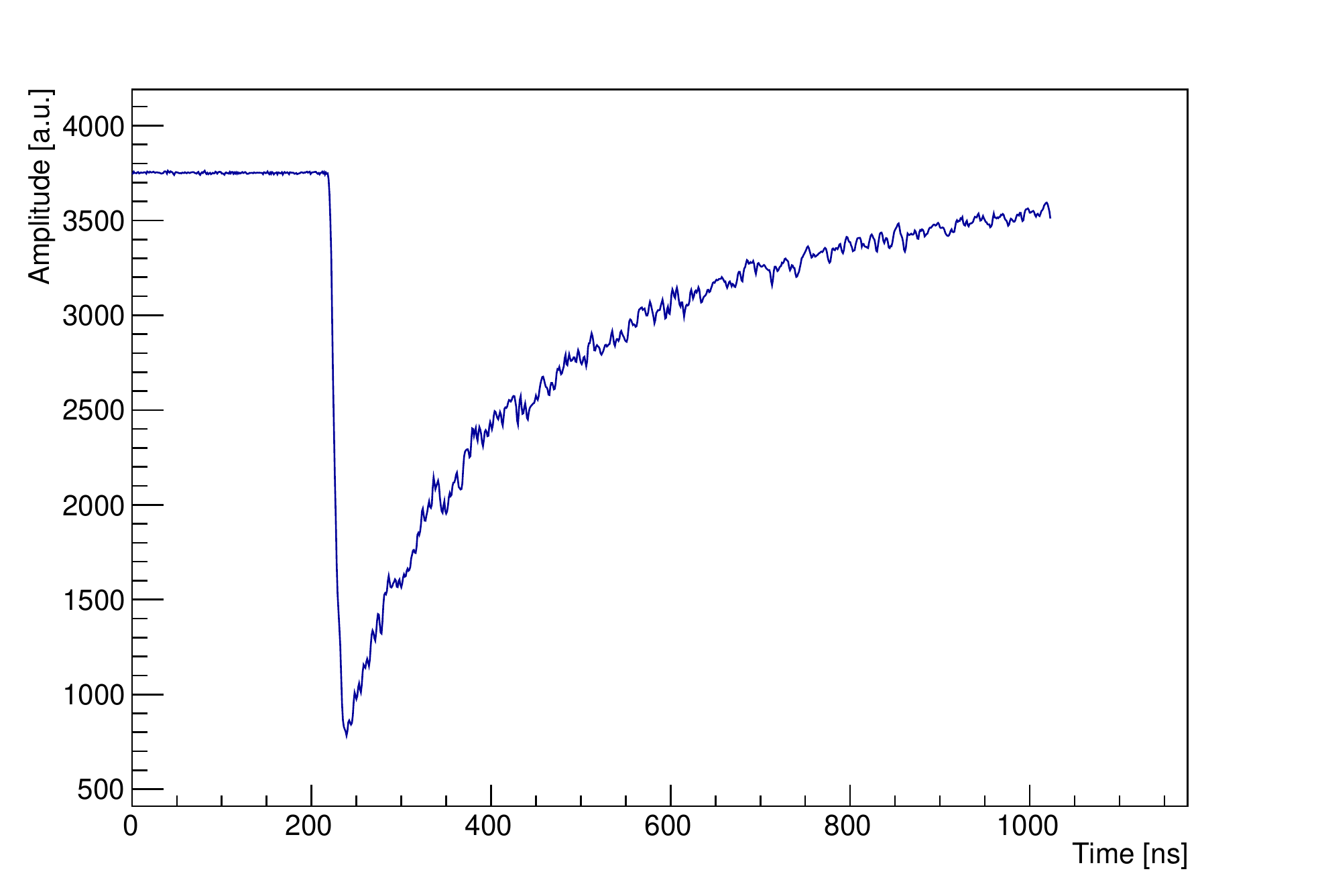}
\par\end{centering}
\caption{\label{fig:BGO signal}An example of a digitised PADME ECal SU signal.}
\end{figure}

SUs signals are digitised by means of CAEN V1742 boards. Each board
is equipped with $4$ DRS4 chips, for a total of $32$ channels. The
channels dynamic range is $[-1,0]\,\text{V}$ with a 12-bit precision,
while the number of sampled points is $1024$. These boards can acquire
data at different sampling frequencies: $1\,\text{GS/s}$, $2.5\,\text{GS/s}$
and $5\,\text{GS/s}$. Given the long BGO signals (example in figure
\ref{fig:BGO signal}), the selected one is $1\,\text{GS/s}$.

Temperature variations are extremely important when using BGO, because
its light yield varies as $-0.9\%/\text{\textdegree C}$. To reduce
as much as possible this variability a new air conditioning system
was installed in the BTF. It is foreseen that the temperature will remain within $\pm0.5\,\text{\textdegree C}$ the desired value. In addition, PT1000 are glued on crystals to monitor
their temperature: $24$ SUs have a thermometer on their back, close
to the PMT, while other $16$ have two sensors glued on the side,
at $1/3$ and at $2/3$ of crystal length, to probe any dependency
on the distance from the front face.

\subsection{Scintillating units calibration with $^{22}$Na}

\begin{figure}[p]
\begin{centering}
\includegraphics[scale=0.35]{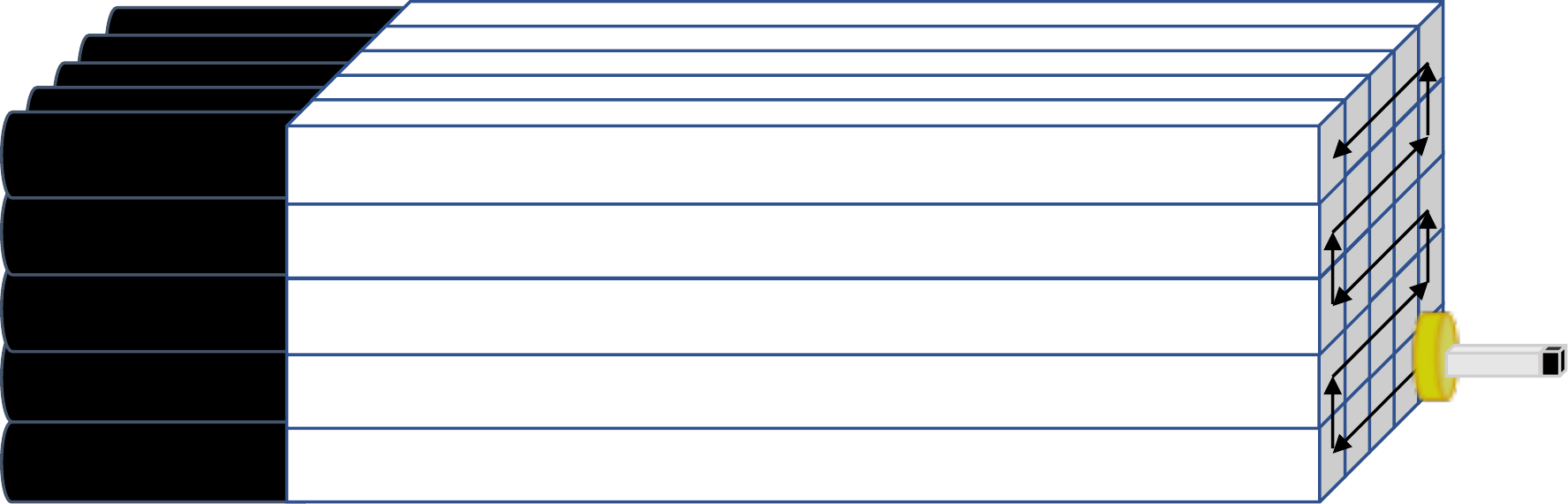}
\par\end{centering}
\caption{\label{fig:Na-22 setup}Scheme of the setup for the SUs calibration
with the $^{22}$Na source. The ECal SUs are on the left, on the right
there is the trigger: a LYSO read by a SiPM. The yellow disk between
the two, encloses the source.}
\end{figure}

Before inserting the SUs in the ECal, each of them has been calibrated
by means of a $^{22}$Na source. The reason for the choice of this isotope is to
exploit its back-to-back emission of $2$ $511\,\text{keV}$ $\gamma$s
coming from an $e^{+}/e^{-}$ annihilation. The calibration setup
consists of a $5\times5$ SUs matrix in front of which the source
is moved by a couple of step motors. The trigger signal is provided
by a $3\times3\times20\,\text{mm}^{3}$ LYSO crystal read by a SiPM,
positioned on the other side of the source with respect to the SU:
a BGO window is acquired every time that the LYSO signal exceeds a
threshold. In figure \ref{fig:Na-22 setup} a scheme of the setup
is presented.

\begin{figure}[p]
\begin{centering}
\includegraphics[scale=0.35]{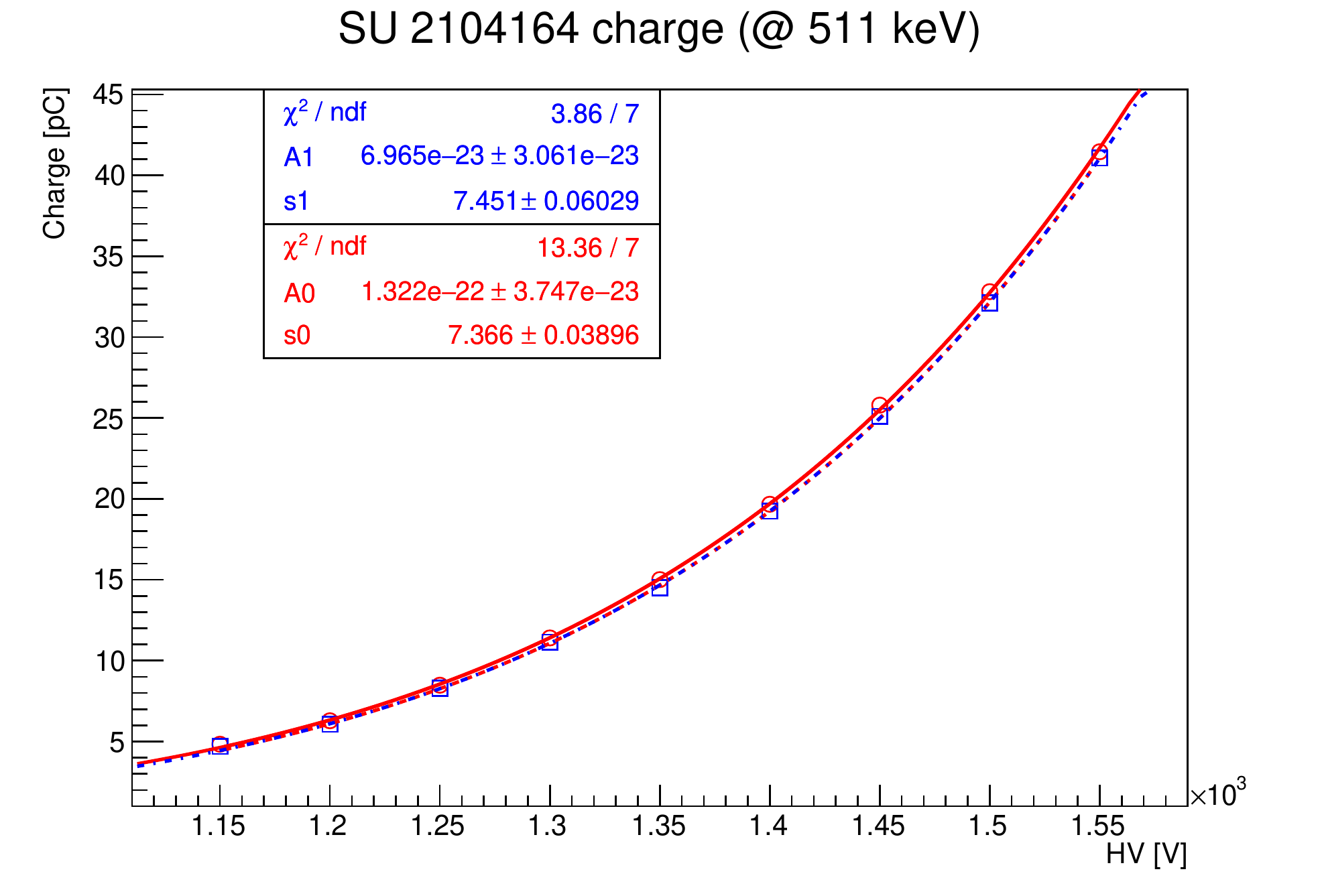}
\par\end{centering}
\caption{\label{fig:SU gain}Charge corresponding to a $511\,\text{keV}$ photon 
from the $^{22}$Na source as a function
of the HV for two measurements performed on the same ECal SU. Only statistical
errors are shown.}
\end{figure}

To construct the charge vs HV curve, all the SUs were measured
at $10$ different voltages in the interval $[1100,1550]\,\text{V}$, in
steps of $50\,\text{V}$, taking about $6000$ events per voltage.
For $135$ SUs, this measurement was repeated twice, to evaluate the stability
of the obtained results. Each set of data is fitted with the function 
$Q=A\cdot V^{s}$ ($Q$ is the charge, $V$ is the HV
and $A$ and\textbf{ $s$} are the fit parameters), that allows to
get the HV value for the desired gain. An example for a single SU,
with the repeated measurement and fit on both groups of data, is given
in figure \ref{fig:SU gain}.

\begin{figure}[p]
\begin{centering}
\includegraphics[scale=0.35]{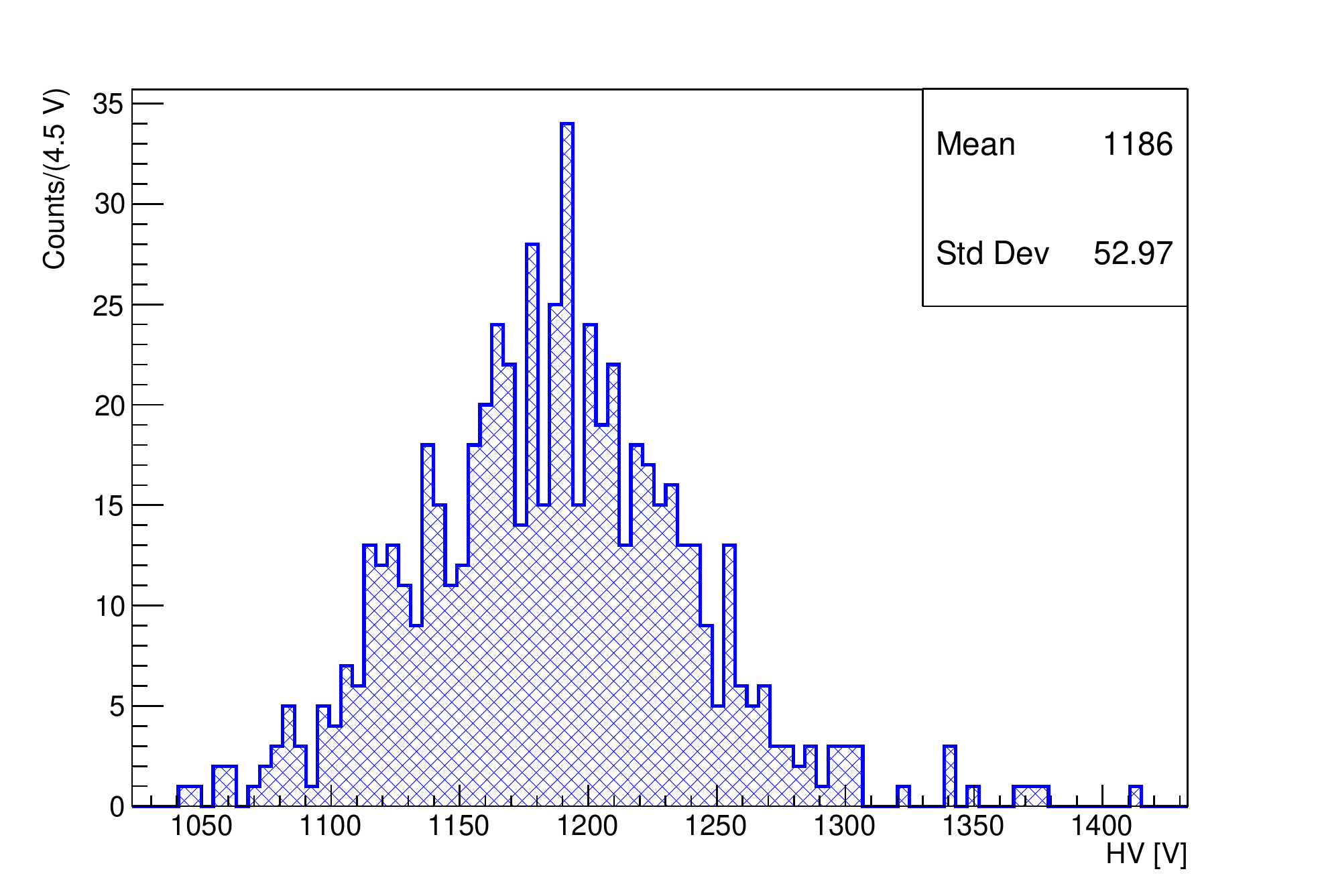}
\par\end{centering}
\caption{\label{fig:HV distr}HV distribution for an equalisation of the ECal
SUs at $15.3\,\text{pC/MeV}$.}
\end{figure}

The HV distribution obtained for all the ECal SUs, when requiring
a gain of $15.3\,\text{pC/MeV}$, is shown in figure \ref{fig:HV distr}.
It presents an average value of $1186\,\text{V}$ with a standard deviation
of $53\,\text{V}$, meaning that at this gain the typical voltage
is well below the maximum safe value of $1700\,\text{V}$ and that
there is a low variability between the units.
This gain is chosen because it is at the centre of the linearity range
and leaves room for signals up to $1\,\text{GeV}$ to be fully contained in the DAQ
without saturating. This is necessary due to the possibility of having events with more than one 
photon in the same SU. In these cases the pulse amplitude from 
each photon may sum to a value larger than the amplitude given by a 
single particle of the maximum energy.

\subsection{Cosmic rays in the electromagnetic calorimeter}

\begin{figure}
\begin{centering}
\includegraphics[scale=0.285]{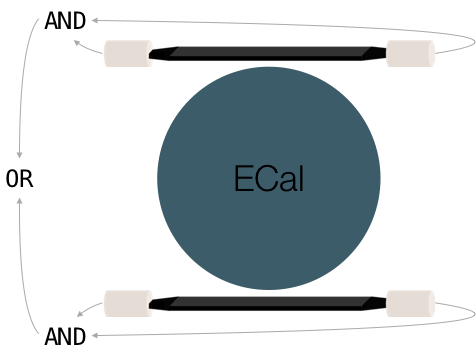}
\par\end{centering}
\caption{\label{fig:CRs trigger}The PADME ECal CRs trigger structure and logic.
Two scintillating paddles, one above and one below the ECal, each read by a
pair of PMTs. The PMTs on the same paddle are in logic AND. 
The logic OR of these two ANDs gives the trigger signal.}
\end{figure}

To study and improve the ECal performances, also Cosmic Rays (CRs)
are exploited. In figure \ref{fig:CRs trigger} the CRs trigger structure
and logic are shown. It consists of two scintillating paddles, one
above and one below the ECal, each one read by two PMTs, one per each
side of the paddle. PMTs on the same paddle are in logic AND. The logic OR of these two ANDs gives the trigger signal. The OR is needed to maximise
the trigger rate and to acquire also cosmics that cross the ECal diagonally, 
which pass through a single paddle.

\begin{figure}
\begin{centering}
\includegraphics[scale=0.35]{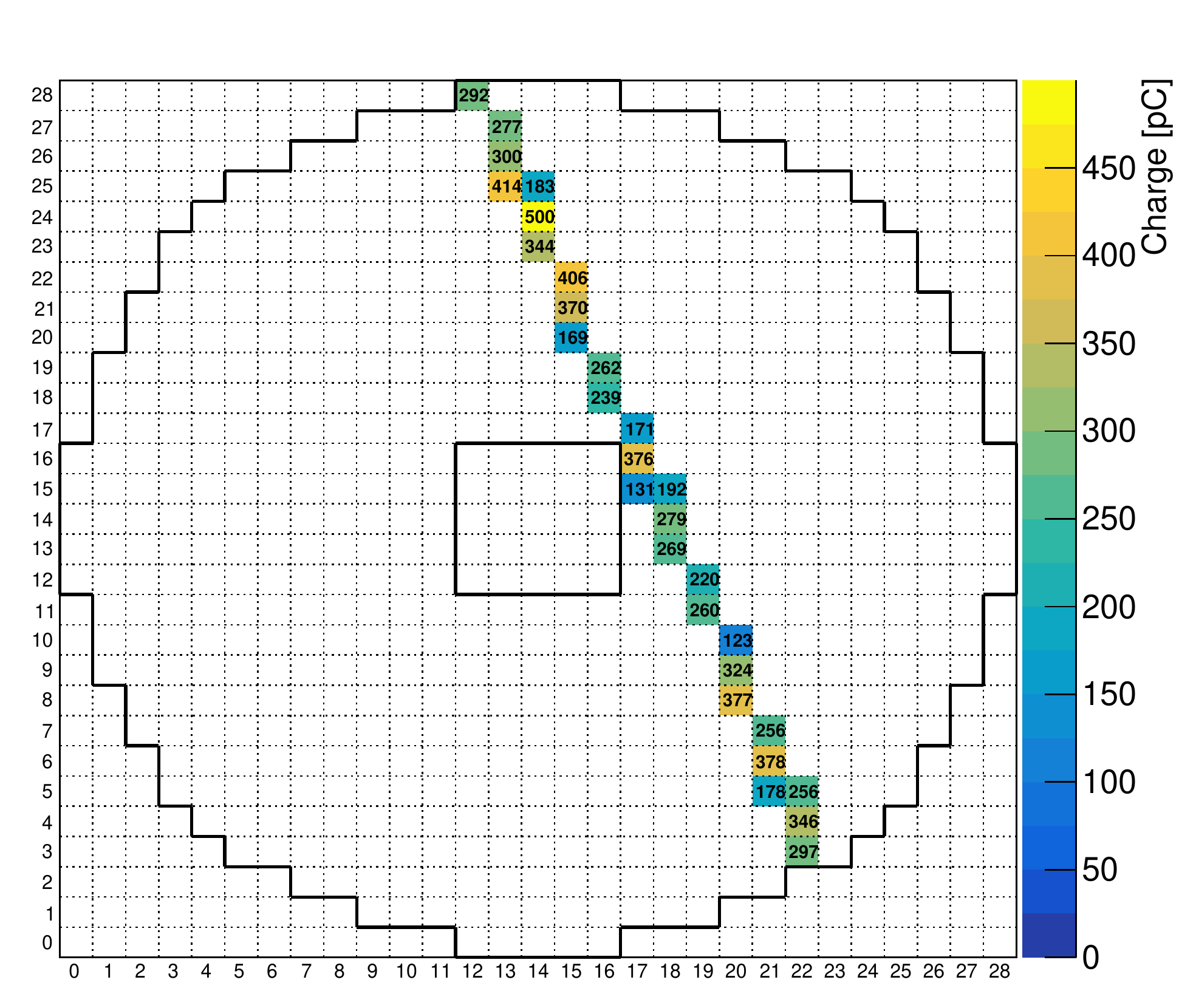}
\par\end{centering}
\caption{\label{fig:CR track}Example of a CR track in the ECal. The colour
scale indicates the charge collected by each SU.}
\end{figure}

In figure \ref{fig:CR track} an example of a CR passing through the
ECal is shown. The color scale, as the number in the box, indicates
the charge integral of the SU pulse in pC.

Using CRs it is possible to evaluate the gain homogeneity between
the units and improve the ECal energy resolution \cite{ECal}. The
selection of CRs vertically passing through a crystal ensures a smaller
spread of the crossing path and, hence, of the released energy. Starting
from this and knowing that the average amount of released energy is
always the same, it is possible to equalise the charge spectra of
the SUs.

Contemporaneously, with the vertical CRs selection, it is also possible
to study the ECal channels efficiency by checking how many times a
CR is seen by a unit when it is seen by the SU above and the SU below
\cite{ECal}.

\subsection{Beam test on a calorimeter prototype}

A $5\times5$ matrix of pre-production SUs was measured in a test
beam at the BTF \cite{ECAL prototype}. The results obtained show
an energy resolution compatible with the one from the L3 experiment \cite{L3}. In
fact the fit to the relative energy resolution, with the function
$\frac{\sigma(E)}{E}=\frac{a}{\sqrt{E[\text{GeV}]}}\oplus\frac{b}{E[\text{GeV}]}\oplus c$,
gives the following parameters: $a=2.0\%$, $b=0.003\%$ and $c=1.1\%$.
Consequently the target energy resolution of $2\%/\sqrt{E}$ has been achieved.

In the same test beam also the linearity of the charge as a function
of the beam energy has been studied and resulted to remain within the $2\%$
up to $1\,\text{GeV}$.

Preliminary studies, using the BTF positron beam directly on the ECal,
indicate that it shows an energy resolution similar or better with
respect to the one obtained with the prototype \cite{ECal}. 

\section{Conclusions}

The PADME experiment main purpose is the search for a dark photon ($A'$) that decays
into dark matter particles, looking for the reaction $e^{+}\,e^{-}\rightarrow A'\,\gamma$.
In this context the electromagnetic calorimeter of the experiment
plays a fundamental role, allowing the measurement of the final state
photon four-momentum and, hence, the $A'$ boson mass, as the reaction missing
mass.
In this paper we describe the calorimeter with the solutions adopted during
its construction aimed to improve its performance, like the scintillating
material choice, the presence of black Tedlar to reduce the light
crosstalk, the absence of a metal structure for the calorimeter and
the usage of PT1000 sensors for temperature monitoring.

In addition the calorimeter cosmic rays trigger is described, together
with the possible studies that can be performed by exploiting this
information.

As a last argument, the results obtained with a calorimeter prototype
are summarised, showing how the observed performance are in line
with design ones.

\end{document}